\documentstyle[11pt,fleqn]{article}
\input epsf

\begin{document}
%\begin{titlepage}
\vspace*{-60pt}
\begin{flushright}
{\small 
SUSSEX-AST 98/1-1 \\
astro-ph/9801148}\\
\end{flushright}
\vspace*{10pt}
\begin{center}
\Large {\bf INFLATION AND THE COSMIC MICROWAVE BACKGROUND}
\vspace{.8cm}
\normalsize

Andrew R.\ Liddle\\
{\it Astronomy Centre, University of Sussex\\ Falmer, Brighton BN1  
9QH, United Kingdom.} 
\date{\today} 
\end{center}
\begin{abstract} 
I give a status report and outlook concerning the use of the cosmic microwave 
background anisotropies to constrain the inflationary cosmology, and stress 
its crucial role as an underlying paradigm for the estimation of cosmological 
parameters.
\end{abstract}
%\end{titlepage}

\section{Introduction}

For a long time now, inflation has been the leading paradigm for the origin 
of cosmological structures. This is largely due to its continuing success in 
confrontation with a wide range of observations, but also due in part to its 
theoretical simplicity compared to rivals such as cosmic strings, both in 
terms of making predictions for the perturbations and in the form (gaussian 
and adiabatic) of perturbations generated.

The interaction between observations and theoretical modelling of inflation 
plays a two-fold role in cosmology. The most-emphasized role is the possible 
use of observations, especially of cosmic microwave background anisotropies, 
to support or rule out the inflationary paradigm as the source of structures 
(see Liddle \& Lyth 1993 for a review). 
Such a program may well offer the first glimpses of possible physics at very 
high energies. Much less has been said about the second role of inflation --- 
that in providing a simple framework for structure formation, it is crucial 
in enabling the high-accuracy determination of more mundane cosmological 
parameters such as the Hubble constant and the density parameter. I shall 
focus particularly on this aspect towards the end of this review.

\section{Inflation}

Inflation is defined as any epoch of the Universe's history during which 
the scale factor $a(t)$ is accelerating. In fact, we have yet to prove 
conclusively that this is not happening at the present epoch, but here I am 
interested in whether such a behaviour might have happened in the Universe's 
distant past. I actually prefer to use the Hubble parameter $H = \dot{a}/a$ 
to write this in a somewhat different way
\begin{equation}
\ddot{a} > 0 \quad \Longleftrightarrow \quad
	\frac{d}{dt} \left( \frac{H^{-1}}{a} \right) < 0 \,,
\end{equation}
where dots are of course time derivatives.
The quantity $H^{-1}$ is the Hubble length, the principal characteristic 
scale of an expanding Universe, and dividing it by $a$ switches to comoving 
coordinates, i.e.~the behaviour of relative to the expansion. In words, 
inflation is precisely the condition that the comoving Hubble length is 
decreasing with time. Since in comoving units all the objects just remain 
where they are as the Universe expands, and since at any epoch the Hubble 
length is a good estimate 
of how far light can travel during that epoch, this is telling us that 
inflation acts like a zoom lens, focussing in on an ever-tinier part of the 
initial region. That's why, for example, inflation can solve the flatness 
problem; even if curvature is important initially, when we zoom in on a tiny 
region the curvature becomes negligible, and enough `zooming' can more than 
compensate for the subsequent increase of the comoving Hubble length after 
inflation ends.

Inflation's ability to generate large-scale perturbations is all down to the 
behaviour of the Hubble length, because it means that a given length scale 
may start well inside the Hubble radius, but finish up well outside it. Any 
irregularities existing at that time become `frozen-in', unable to evolve.
If we lived in a purely classical world, this would lead to a perfectly 
homogeneous Universe, simply because the assumption that the initial energy 
density be finite requires that there cannot be irregularities down to 
arbitrarily small scales.

Fortunately, we live not in a classical Universe but a quantum one, and the 
uncertainty principle implacably opposes inflation's attempts to make a 
perfectly smooth Universe. More so, it does this in a manner which is readily 
predictable for a given inflationary model. It turns out that quantum 
fluctuations give rise to two types of perturbations in the Universe, density 
perturbations and gravitational waves. These both take on a classical 
character once they are on scales well in excess of the Hubble length.

In order to give definite predictions for these, it is necessary to have a 
definite model for the inflationary expansion rather than just the 
hand-waving sketch I've given above. Here I want to stress that inflationary 
model-building has become quite a complicated and sophisticated occupation, 
and in presentations like this one necessarily gives an oversimplified 
picture. Bear that in mind, and I'll also remind you of it later. 

\begin{figure}[t]
\centering 
\leavevmode\epsfysize=5cm \epsfbox{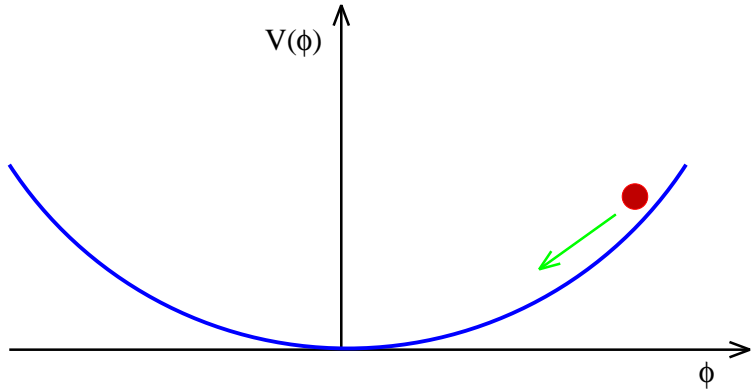}\\ 
\caption[scalpot]{\label{scalpot} A schematic scalar field potential.} 
\end{figure} 
 
The acceleration equation
\begin{equation}
\frac{\ddot{a}}{a} = - \frac{4\pi G}{3} \left( \rho + 3p \right) \,,
\end{equation}
immediately tells us that we're going to need something a little out of the 
ordinary, namely $p < - \rho/3$. The standard way of achieving this uses 
a scalar field, the sort of thing which crops up all over the place in modern 
particle physics theories, especially when symmetry breaking is under 
consideration. The standard simplified picture, which does in fact cover a 
large fraction of the currently popular models, is that there is only one 
scalar field, and that it evolves classically by slow-rolling down a 
self-interaction potential $V(\phi)$ such as that shown in 
Figure~\ref{scalpot}. 
Either of these assumptions can be altered, giving more complicated models, 
but the bulk of my discussion will stay with the simplest case.

Such a homogeneous scalar field has effective energy density and pressure
\begin{equation}
\rho_\phi = \frac{1}{2} \dot{\phi}^2 + V(\phi) \quad ; \quad 
	p_\phi = \frac{1}{2} \dot{\phi}^2 - V(\phi) \,,
\end{equation}
and so the condition for inflation is satisfied as long as the potential 
dominates over the kinetic term. This will clearly happen if the potential is 
sufficiently flat; in fact, the flatness condition is very weak and so 
inflation is quite generic.

In principle, $V(\phi)$ is predictable from fundamental theories of physics. 
In practice there is as yet no clear guidance, and instead we treat it as a 
free function to be constrained by observations.

\section{Describing the perturbations}

If the inflationary expansion is rapid enough, physical conditions during 
inflation will change little between the origin of perturbations on the 
largest interesting scales and the smallest, so as a rough rule of thumb we 
expect that the perturbations will be nearly scale-invariant. For a long 
time, the perturbations were indeed taken to have that form. However, more 
recently the observations have reached such quality that the scale-invariance 
approximation is no longer an adequate description, and we have to do better. 
This is in no way a set-back (or climb-down) for inflation --- it is an 
impressive success that the theory has done so well that we now regard small 
corrections to the initial picture as significant and observationally 
testable.

The breaking of scale-invariance will depend on the form of the potential 
$V(\phi)$ (it being the only input information), so we quantify this by 
defining two slow-roll parameters (Liddle \& Lyth 1992)
\begin{equation}
\epsilon \equiv \frac{m_{{\rm Pl}}^2}{16 \pi} \left( \frac{V'}{V} \right)^2
	\quad ; \quad \eta \equiv \frac{m_{{\rm Pl}}^2}{8 \pi} 
	\frac{V''}{V}  \,,
\end{equation}
where prime means a derivative wrt $\phi$, and $m_{{\rm Pl}} = G^{-1/2}$ is 
the Planck mass. Inflation requires that both these be less than one, in 
order to maintain dominance of the potential. The first measures the slope of 
the potential, the second its curvature.

The terminology I will use isn't very important; density perturbations are 
specified by $\delta_{{\rm H}}(k)$ and gravitational waves by $A_{{\rm 
G}}(k)$ where $k$ is the comoving wavenumber. For scale-invariant spectra 
these are both independent of $k$; the power spectrum $P(k) \propto k 
\delta_{{\rm H}}^2$. Under the inflationary paradigm, these spectra are 
responsible for all the observed structures, with gravitational waves at best 
only significantly influencing large-angle microwave background anisotropies.

The formulation I'll describe is based on a perturbative approach, where we 
expand the (log of the) spectra in terms of log(wavenumber) about some scale 
$k_*$:
\begin{equation}
\label{taylor}
\ln \delta_{{\rm H}}^2(k) = \ln \delta_{{\rm H}}^2(k_*) + (n_* -1)
	\ln \frac{k}{k_*} + \frac{1}{2} \left. \frac{dn}{d\ln k} \right|_*
	\, \ln^2 \frac{k}{k_*} + \cdots \,,
\end{equation}
and truncate at some level. For example

\indent \indent {\bf First term:} Harrison--Zel'dovich spectrum (constant 
$\delta_{{\rm H}}$).

\indent \indent {\bf Second term:} Power-law spectrum with spectral index 
$n$.

\indent \indent {\bf Third term:} Includes scale-dependence of the spectral 
index.

Current observations require at least the second term and normally people 
stop there. I'll return to the third one later. The second term gives the 
power-law approximation, and the observables can be very nicely computed in 
terms of the slow-roll parameters (Liddle \& Lyth 1992):
\begin{equation}
n = 1 - 6\epsilon + 2 \eta \quad ; \quad n_{{\rm G}} = -2\epsilon 
	\quad ; \quad
\frac{A_{{\rm G}}^2(k_*)}{\delta_{{\rm H}}^2(k_*)} = \epsilon = 
	- \frac{n_{{\rm G}}}{2} \,.
\end{equation}
If slow-roll holds very well ($\epsilon \ll 1$, $|\eta| \ll 1$) we get 
scale-invariant density perturbations and negligible gravitational waves. 
Note also that although $\epsilon$ is positive by definition (the field 
always rolls downhill), $\eta$ can have either sign and $n$ can be greater or 
smaller than one. The final relation, giving the ratio of the spectra in 
terms of $n_{{\rm G}}$, is known as the consistency relation; it 
represents the inevitable `entanglement' of the density perturbations and 
gravitational waves due to their common origin in the single function 
$V(\phi)$. In the unlikely event of it proving testable, it represents a very 
distinctive prediction of inflation.

\section{COBE}

The COBE observations have a very simple interpretation in terms of 
inflation. Because the COBE beam is so wide, it only probes scales larger 
than the horizon at the time of decoupling, so perturbations have not had 
time to evolve and are captured in their primordial form. In particular, the 
observed anisotropies do not depend on cosmological parameters such as the 
hubble constant $h$ and the baryon density $\Omega_{{\rm B}}$.

COBE determines the perturbation amplitude extremely well, to about 10\% 
accuracy, but is unable to distinguish the effect of density perturbations 
and gravitational waves. Its most useful application is to normalize the 
density perturbation spectrum; Bunn et al.~(1996) obtained the result
\begin{equation}
\delta_{{\rm H}} = 1.91 \times 10^{-5} \, \frac{\exp \left[ 1.01(1-n)
	\right]}{\sqrt{1+0.75 r}} \,,
\end{equation}
using techniques described by Bunn \& White (1995, 1997). Here $\delta_{{\rm 
H}}$ is the perturbation amplitude at the present Hubble radius, and $r = 
12.4 
A_{{\rm G}}^2/\delta_{{\rm H}}^2$ approximately measures the relative 
importance of gravitational waves and density perturbations in generating the 
anisotropies. The factor 12.4 comes from analytic evaluation assuming only 
the Sachs--Wolfe effect applies and perfect matter domination at last 
scattering; that the above expression contains the factor 0.75 indicates that 
this approximation fails at the tens of percent level on COBE scales.

COBE fixes the energy scale of inflation as (Bunn et al.~1996)
\begin{equation}
V_*^{1/4} = \left( 6.6 \times 10^{16} \, {\rm GeV} \right) 
	\epsilon^{1/4} \quad \pm 5\% \,,
\end{equation}
where $*$ indicates the value when the observed perturbations were generated. 
For a specific model $\epsilon$ is known and so this can be given exactly. 
Unless $\epsilon$ is tiny, which is perfectly possible in some models, the 
energy scale is around that expected of Grand Unification.

The spectral index $n$ is only very weakly constrained by COBE, due to the 
limited range of scales sampled.

\begin{figure}[t]
\centering 
\leavevmode\epsfysize=10cm \epsfbox{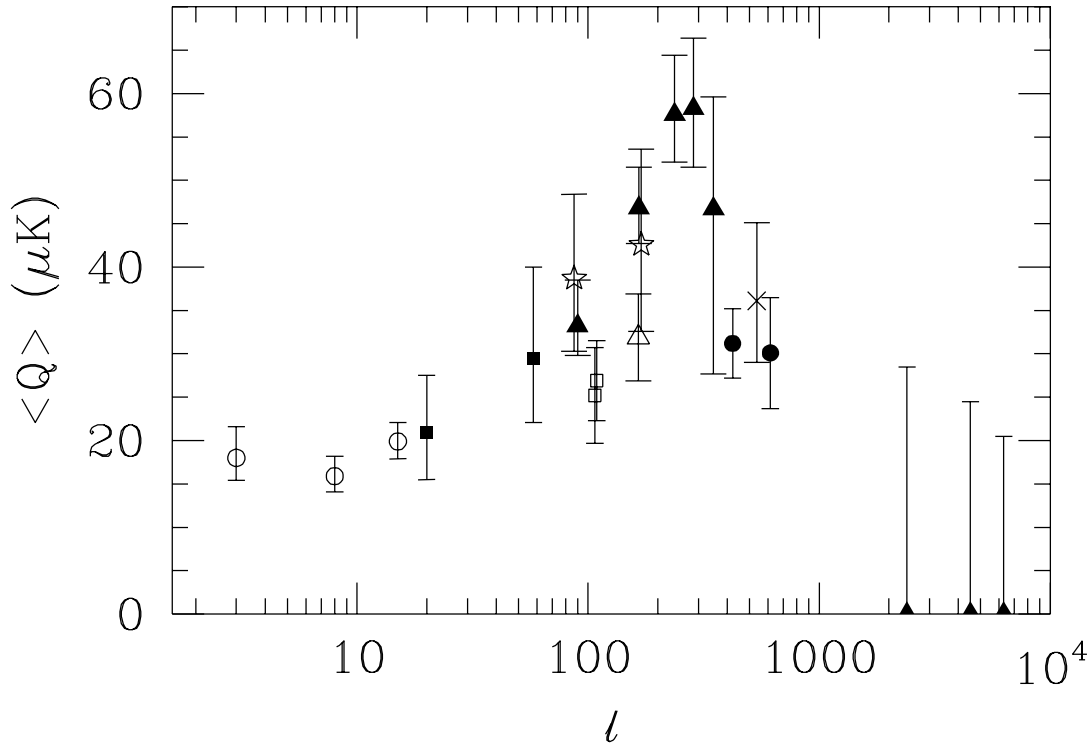}\\ 
\caption[cell]{\label{cell} A compilation of measurements of microwave 
background anisotropies, by Martin White. The detections (1-sigma errors) are 
COBE (3 open circles), Tenerife and BAM (filled squares), Python (2 stars), 
Saskatoon (5 filled triangles), ARGO (2 open squares), MAX (open triangle), 
CAT (2 filed circles) and OVRO (cross). The 95\% confidence upper limits are 
SuZIE, ACTA and Ryle. For a detailed discussion see Anthony Lasenby's 
contribution to these proceedings.} 
\end{figure} 
 
\section{The current compilation}

Nowadays, measured anisotropies go well beyond COBE, with a host of 
experiments reporting detections on a range of angular scales. Figure 
\ref{cell} shows a recent compilation by Martin White. The data are 
encouragingly compatible with inflationary preconceptions, but unfortunately 
at present don't allow us to say much more; the predictions on smaller 
angular scales (larger $\ell$) depend on all the cosmological parameters and 
the observational errors are larger than one desires. The compilation does 
indicate a lower limit on $n$, which is somewhat model-dependent but is 
around $0.75$, and at the moment that is its main inflationary implication.

\section{Solving cosmology ...}

The upcoming launch of the MAP and Planck satellites offers the prospect that 
cosmology could be more or less solved, in the sense that its most crucial 
parameters might be measured to a satisfyingly high accuracy (Jungman et 
al.~1996; Bond et al.~1997; Zaldarriaga et al.~1997). However, there 
are rather a lot of parameters which might come into play. Most attention has 
been focussed on the cosmological parameters, such as the Hubble constant 
$h$, total matter density $\Omega_0$, baryon density $\Omega_{{\rm B}}$, 
cosmological constant $\Omega_\Lambda$, hot dark matter density $\Omega_{{\rm 
hdm}}$ 
and the optical depth to the last-scattering surface $\tau_{{\rm c}}$. Some 
of these might be fixed by assumption, but as a very minimum $h$, 
$\Omega_{{\rm B}}$ and $\tau_{{\rm c}}$ must be determined observationally.

However, in this article I want to turn attention to the inflationary input. 
As I stressed at the beginning, this is absolutely crucial as a paradigm in 
cosmology. If we don't know the initial perturbation spectra, we have no 
chance of interpreting observed microwave anisotropies in terms of the 
cosmological parameters. This is potentially a very serious problem, since 
without guidance the perturbations could be free functions which need not be 
simple.

Fortunately, most models of inflation give very simple predictions, and are 
summarized in only two or three parameters, namely $\delta_{{\rm H}}$, $n$ 
and perhaps $r$. However, scale-dependence of $n$ is possible, and in 
accordance with the Taylor expansion of Eq.~(\ref{taylor}) it can be 
incorporated by inclusion of derivatives $dn/d\ln k$, $d^2 n/d \ln k^2$ and 
so on.
These represent extra parameters which must be determined from observations. 

If extra parameters are introduced, then the determination of {\em all} 
parameters will deteriorate. We have estimated the extent of this 
deterioration (Copeland et al.~1997) for a configuration of the Planck 
satellite including polarized detectors. It is shown in Table \ref{parest}.

\begin{table}[t]
\centering
\caption{\label{parest} Estimated parameter errors (one-sigma) for the 
Standard CDM model, as extra scale-dependence is introduced.} 
\begin{tabular}{llll} 
\hline
Parameter &\multicolumn{3}{c}{Planck 140 GHz channel} \\ 
  & \multicolumn{3}{c}{with polarization} \\ 
\hline
$\delta \Omega_{{\rm b}} h^2 /\Omega_{{\rm b}} h^2$ & $0.007$ & $0.009$ & 
	$ 0.01$ \\
$\delta \Omega_{{\rm cdm}} h^2 /h^2$ & $0.02$ & $0.02$ & $ 0.02$ \\
$\delta \Omega_\Lambda h^2 /h^2$ & $0.04$ & $0.05$ & $ 0.05$ \\
$\delta \tau_{{\rm C}}$ & $0.0006$ & $0.0006$ & $0.0006$ \\ 
\\
$\delta n$ & $0.004$ & $0.04$ & $0.14$ \\
$\delta r$ & $0.04$ & $0.05$ & $0.05$ \\
$dn/d\ln k$ & $-$ & $0.006$ & $0.04$ \\
$d^{2}n/d(\ln k)^2$ & $-$ & $-$ & $0.005$ \\
\hline
\end{tabular}
\end{table}

We see that permitting scale dependence of $n$ primarily only affects the 
determination of $n$, and not the other cosmological parameters, which is an 
encouraging conclusion. It suggests that if the modelling of inflation turns 
out to be over-simplified, the influence on parameter determination will not 
be too great. On the other hand, it is clear that $n$ cannot be measured as 
well as has been claimed (e.g.~Bond et al.~1997; Zaldarriaga et al.~1997) 
unless it is {\em assumed} that the spectrum is a perfect power-law.

\section{Inflationary complications}

Now, as promised, I turn to the question of possible complications to the 
inflationary modelling. First of all, one can ask whether we are in a 
position to compute the perturbations at the 1\% or so accuracy level 
demanded by MAP and Planck. Finally, the answer to this appears to be `yes'; 
the last problem (gravitational waves in open Universe models --- see below) 
has been solved during the last year and there are no existing models in 
which the perturbations cannot be computed, at least through numerical 
integration of the relevant mode equations, to the required accuracy. 

\subsection{Single-field models}

So far we've been sticking to models with a single scalar field, with 
$V(\phi)$ kept as a free function. Normally the slow-roll approximation gives 
a very accurate analytic result (Grivell \& Liddle 1996).
However, a sufficiently complicated potential may lead to a failure of 
slow-roll severe enough that the slow-roll approximation is not good enough 
(Wang et al.~1997), and the numerical results are required. However, we have 
found (Copeland et al.~1998) that this need not be a bad thing; in 
particular, the near-failure of slow-roll makes it far more likely that the 
scale-dependence of $n$ is observable, which allows one to determine more 
information about the inflaton potential than would otherwise have been 
available. Further, it would take a bizarre conspiracy for the slow-roll 
approximation to be failing, but yet for us to be oblivious to this. If 
things are going wrong to the extent that numerical computation is needed, we 
will know it.

The main worry in fact for the single-field models described so far is that 
many models [particularly those constructed under the currently-popular 
hybrid inflation strategy (see Lyth 1996 for a review)] predict a negligible 
level of gravitational waves. This limits seriously the amount of information 
that can be inferred about $V(\phi)$, which cannot even be determined 
uniquely if the gravitational waves cannot be detected.

\subsection{Multi-field models?}

It is certainly possible that more than one scalar field can be dynamically 
important, and this can lead to a range of new phenomena. The perturbations 
may have an isocurvature component as well as the usual adiabatic one, and 
may even be non-gaussian. The perturbations can still be computed accurately, 
but now only on a model-by-model basis rather than via an all-encompassing 
formalism like that I've demonstrated for single-field models. This makes it 
much harder to `guess' viable models from the observations.

However, calculational complications aside, a specific model of this type is 
as easy to exclude using observations as a single-field model.

\subsection{Open inflation models?}

Open inflation models are an unfortunate late addition into the inflationary 
model zoo. Although dating back almost to the beginnings of inflation (Gott 
1982), it is only relatively recently that they have been appreciated as a 
serious model. They rely on quantum tunnelling to generate an open Universe 
within an inflationary sea.

For a long time the perturbations in these models could not be computed 
(especially the gravitational waves), but finally the technology is in place 
(Sasaki et al.~1997; Bucher \& Cohn 1997), and predictions can be obtained 
from them as readily as the more conventional models.

Mathematically they are much nastier models (for example the mode functions 
on hyperbolic geometry are very unpleasant indeed) than ones giving a flat 
spatial geometry. Fortunately, the spatial geometry is very readily 
measurable, even before the satellites go up, and hopefully these models will 
soon be consigned to the dustbin.

\subsection{Inflation not correct?}

If inflation is not in fact the correct theory for the origin of 
perturbations, this should be obvious from the observations. The inflationary 
prediction of passive, super-horizon perturbations is very distinctive, and 
leads to the familiar oscillations in the radiation angular power spectrum. 
If inflation is correct the observations should highly overdetermine the 
various parameters of the big-bang model, reassuring us that we are on the 
right track.

But here is a good point to stress once more the importance of inflation as a 
paradigm for the initial conditions. For example, it seems rather unlikely 
that predictions at the one percent level could come from the rival 
topological defect scenario in the forseable future, due to the horrendous 
non-linearities involved in the computations. Without accurate theoretical 
predictions for the anisotropies, huge extra uncertainties enter the 
cosmological parameter estimation game and severely dent one's ability to 
measure any of them.

\section{Conclusions}

My main point of emphasis has been to stress that the inflationary paradigm 
is a crucial underpinning of attempts to measure cosmological parameters from 
the microwave background. It provides a framework in which accurate 
predictions can readily be made, enabling the maximum to be squeezed out of 
quality observations. We are fortunate indeed that not only is the model 
theoretically appealing, but that it also stands in excellent shape when 
confronted with present data. We can only hope that in ten years time, when 
the large zoo of inflation models have been confronted with the new 
observations, that things may still look so good.

\section*{Acknowledgments}

I was supported by the Royal Society. Thanks to Ted Bunn, Ed Copeland, Ian 
Grivell, Rocky Kolb, David Lyth and Martin White for their collaboration in 
parts of the work described here, and to Martin White for providing the data 
compilation shown in Figure~\ref{cell}.

\section*{References}

\begin{description}
\item Bond, J.R., Efstathiou, G. and Tegmark, M., 1997, Mon. Not.
	R. Astron. Soc. {\bf 291}, L33
\item Bucher, M. and Cohn, J.D., 1997, Phys. Rev. D {\bf 55}, 7461
\item Bunn, E.F., Liddle, A.R. and White, M., 1996, Phys. Rev. D
	{\bf 54}, 5917R
\item Copeland, E.J., Grivell, I.J. and Liddle, A.R., 1997,
	Sussex preprint {\tt astro-ph/9712028}
\item Copeland, E.J., Grivell, I.J., Kolb, E.W. and Liddle, 
	A.R., 1998, Sussex preprint 
\item Gott, J.R., 1982, Nature {\bf 295}, 304
\item Grivell, I.J. and Liddle, A. R., 1996, Phys. Rev. D {\bf 54}, 
	7191
\item Jungman, G., Kamionkowski, M., Kosowsky, A. and 
	Spergel, D.N., 1996, Phys. Rev D {\bf 55}, 7368
\item Liddle, A.R. and Lyth, D.H., 1992, Phys. Lett. B 
	{\bf 291}, 391
\item Liddle, A.R. and Lyth, D.H., 1993, Phys. Rep. {\bf 231}, 1
\item Lyth, D.H., 1996, Lancaster preprint {\tt hep-ph/9609431}
\item Sasaki, M., Tanaka, T and Yakushige, Y., 1997, Phys. Rev. 
	D {\bf 56}, 616
\item Wang, L., Mukhanov, V.F. and Steinhardt, P.J., 1997,
	Phys. Lett. B {\bf 414}, 18
\item Zaldarriaga, M., Spergel, D. N. and Seljak, U., 1997,
	Astrophys. J. {\bf 488}, 1
\end{description}
 
\end{document}